\documentclass[prb,twocolumn,superscriptaddress,showpacs]{revtex4-1}

\usepackage{graphics}           
\usepackage{bm}                 
\usepackage{amsmath}            
\usepackage{verbatim}           
\usepackage{amsfonts,amssymb,latexsym,amscd}
\usepackage{braket}             
\usepackage{color}

\bibpunct{[}{]}{,}{n}{}{}     

\begin{document}

\title{Estimating the Entanglement Negativity from low-order moments of the partially 
transposed density matrix}


\author{H.A.~Carteret}
\noaffiliation
\affiliation{
         Department of Physics and Astronomy, University of Waterloo,}
\affiliation{
         Institute for Quantum Computing, University of Waterloo, \\
         200 University Avenue West, Waterloo, Ontario, N2L 3G1, Canada}

\date{February 6, 2017}

\begin{abstract}
We show how to find families of infima and suprema for the entanglement negativity using 
only a few, low-order moments of the partially transposed density matrix $\rho^{T_2}.$ 
These moments can be measured using the multi-copy quantum circuits previously 
given by the author, which define a set of multi-copy expectation values and thus can be 
used with the replica trick. As such, these bounds are suitable for use with Quantum Monte 
Carlo methods, and the lower order versions of the estimates may be experimentally 
accessible for some systems. 
Using more moments for higher-order versions of these methods will produce tighter 
estimates, unless and until statistical noise causes the measurement of the highest order 
moment to fail. Should this happen, the data from lower order moments can still be used 
for lower-order estimates. 
\end{abstract}



\maketitle


\section{Introduction}

It was recently shown that the replica trick \cite{EEandQFT} can be used to measure 
the R{\'e}nyi entropies for single subsystems of a lattice in a globally pure state 
\cite{RenyiMelko}. 
This allows the quantification of the entanglement across the cut but only if the state of 
the entire system is pure.  As such, it cannot tell us anything about the entanglement 
between two subsystems, or what happens at non-zero temperatures. Moreover there have also 
been developments in adapting this method for use in experimental systems 
\cite{NatureTwins}, as well as methods \cite{twoPhoton1,twoQubit2} 
based on the circuits for measuring the moments of the partially transposed density 
matrix developed by the author in \cite{PT}. These expectation values have already been 
studied as individual moments using Quantum Monte Carlo in \cite{Lauchli3}, and the reader 
is referred to that paper as well as \cite{RenyiMelko,genericRenyi,pathIntegralRenyi} 
for the details of sampling methods.

The problem of determining whether a state is bipartite entangled is known to be NP-hard 
in the size of the subsystems\cite{GurvitsSTOC}, see also \cite{Gharibian,Ioannou}. 
However, the problem of determining whether multiple copies of a state can be ``distilled'' 
\cite{firstDistillation} into a smaller number that are more entangled is believed to be 
a lot more tractable as the subsystems involved become large. This question can be 
addressed by evaluating an entanglement measure \cite{computEnt,negativity1} known as 
the \emph{entanglement negativity}, $\mathcal{N}(\rho).$ 
This is the sum of the moduli of the negative eigenvalues of $\rho^{T_2},$ where 
$\rho^{T_2}$ is the partially transposed density matrix. This can be written in terms of
the one-norm $||X||_1,$ which is defined to be the sum of the moduli of the eigenvalues of 
$X,$ thus 
\begin{equation}\label{linearNegativity}
 \mathcal{N}(\rho)= \frac{1}{2}\left(||\rho^{T_2}||_1-1\right)
\end{equation}
or equivalently, as the log-negativity:
\begin{equation}\label{logNegativity}
 \mathcal{E}(\rho)= \ln(||\rho^{T_2}||_1).
\end{equation}
The entanglement negativity is believed to have a specific operational meaning: given an 
infinite number of copies of the state $\rho,$ what is the asymptotic yield of some 
maximally entangled state that we extract by doing local measurements on all the copies 
\cite{computEnt}?
The entanglement negativity gives an upper bound on the yield for bipartite 
entanglement distillation protocols \cite{computEnt}, which is conjectured to be an equality.

The negativity is also a good choice of entanglement measure because it is defined for 
all bipartite states, including mixed ones, as it is relatively simple to calculate; 
no convex roofs over ensemble decompositions of the density matrix are required. 
Thus a general method for evaluating the negativity would allow us to probe the entanglement 
properties of finite temperature systems without modification.
Indeed, many other natural choices of entanglement and correlation measures 
including the entanglement cost, relative entropy of entanglement and the quantum discord 
are NP-hard/NP-complete \cite{discordNPcomplete} in the subsystem size. 
(This reference also contains a more complete list of computationally hard measures; note 
that the negativity is not on that list.) 

The setting we have in mind is one in which we have two subsystems of interest, as opposed 
to the single subsystem of interest that is usually considered in calculations of the 
R{\'e}nyi entropy. 
Rather than look at the behaviour of our entanglement measure only in the limit as the 
subsystems of interest become large, instead we would also like to explore how the entanglement 
between the two subsystems scales as the subsystems are moved apart. 
This will allow us to detect the presence of long range entanglement and other corrections 
to area laws which cannot be detected using two-point functions alone. 

Some form of long-range entanglement is believed to be related to topological quantum order 
\cite{XGWenBook1,KimLongRangeNec}, though whether or not the negativity captures the relevant 
type of entanglement seems to depend on the system. Various authors have tried to use the 
entanglement negativity as a measure of long-range entanglement in systems that are tractable 
using analytic methods. Some topologically ordered systems do not exhibit long-range negativity 
\cite{CastelnovoToric,LeeVidal}, 
but some other systems do
\cite{EdgeNegaChernSimons,N-2dHarmonicCorner} 
and where present, there does seem to be a connection with topological quantum order.
Several other authors have already examined the entanglement negativity using a combination 
of analytic and numerical methods; small systems can be handled directly using exact 
diagonalization \cite{ultracold}. 
Larger systems have been studied using a variety of methods. Some authors examine the moments 
of $\rho^{T_2}$ only, or the quotients
\begin{equation}
 R_n(y)=\frac{Tr((\rho_{AB}^{T_B})^n)}{Tr(\rho_{AB}^n))}
\end{equation}
introduced in \cite{NegativityQFT}; see also \cite{FiniteT-N-CFT}. 
This approach was combined with Quantum Monte Carlo methods in \cite{AlbaNegativity,Lauchli3}.
More generally, several authors have aleady used the entanglement negativity in numerical 
studies to probe the entanglement properties of specific phenomena 
\cite{NsepBlocksCritChain,N-RandomSpinChains,N-universalityLMG,NegativityKondoChain,Nega2dFreeLattice}.

There have also been attempts to use such an analytical extrapolation, in 
\cite{disjCFTnumExtrap}, though as yet these methods are uncontrolled.
If existence or non-existence proofs are desired, 
such as when considering the ways in which entanglement can be shared between 
multiple systems,
then the extrema are required instead of some kind of maximum entropy estimate. 
(Moreover, it is known that maximum entropy methods must be used with care 
when trying to estimate the distillable entanglement 
\cite{fakeEntanglementH3,EntangUnknownStates,infoVsDistillEnt}.) 

A note on terminology: the word ``negativity'' can have two inequivalent meanings in the 
quantum information literature. One of them is the entanglement negativity defined above; the 
other one refers to situations where quantities that are usually interpreted as probabilities 
can nevertheless take on negative values. (These are sometimes called quasi-probabilities,  
see for example \cite{Hakop,quasiprobQC}.) 
For the rest of this paper, the word ``negativity'' should be interpreted as referring to the 
entanglement negativity only.

\section{Outline of moment problems}

The negativity can be computed exactly for finite-dimensional systems either by exact 
diagonalization of $\rho^{T_2}$ or by first obtaining a complete set of moments  
$\mu_n=\text{Tr}((\rho^{T_2})^n)$ 
and then using these in the Newton-Girard formulae to find the eigenspectrum \cite{PT}. 
(This reference explains the construction in terms of quantum circuits but those can be 
re-interpreted as examples of the multi-copy replica trick in a natural way, by the same 
argument as given in \cite{RenyiMelko}.) 
However, in the context of large system sizes or experimental settings, even finding the 
eigenspectrum has a prohibitive overhead, whether in terms of computational cost or the 
difficulty of precision control of quantum systems.  
A useful method can use at most a few, low-order moments of the partially-transposed 
density matrix $\rho^{T_2},$ which can be written 
\begin{align} 
  \mu_n &= \text{Tr}((\rho^{T_2})^{n}) =\sum_i |\lambda_i|^n, &\text{for} \; n=2m,\label{evenMmts} \\
  \nu_n &= \text{Tr}((\rho^{T_2})^{n}) 
   =\sum_{\lambda_i>0} \lambda_i^n -\sum_{\lambda_i<0} |\lambda_i|^n, &\text{for} \; n=2m-1
\end{align}
Approximate values for these moments can be obtained from a Quantum Monte Carlo routine 
\cite{RenyiMelko,Lauchli3}, using the index contractions given in ref.~\cite{PT}. 
(We use different names for the moment sets to avoid confusion between 
$\nu_1=$Tr$(\rho^{T_2})=1$ and $\mu_1=||\rho^{T_2}||_1,$ 
which is the quantity we are trying to estimate!)  

The moments of the partially transposed density matrix have two distinct analytic 
continuations \cite{CCT-EN-FT}. The analytic continuation of the even-powered moments 
$\widetilde{\mu}(n)$ can be obtained by allowing $n$ to take continuous values in 
\eqref{evenMmts}, and it has the limiting behaviour
\begin{equation}\label{evenContinuation}
 \lim_{n \rightarrow 1} \widetilde{\mu}(n) = ||\rho^{T_2}||_1.
\end{equation}
The odd powered moments $\nu_n$ have their own analytic continuation, which must tend to $1$ 
in the limit as $n\rightarrow 1$ since the partial transpose is a trace-preserving map. 
We will not be using an analytic continuation for the odd moments in this paper, only the 
odd-powered moments $\nu_n$ themselves.

It can be seen by repeated differentiation with respect to $n$ that the even analytic 
continuation $\widetilde{\mu}(n)$ is downward convex. 
Furthermore, if we take the natural logarithms of the even moments $\mu_{n}$ it can be seen 
by repeated differentiation that $\ln(\widetilde{\mu}(n))$ is also downward convex.
Thus if we pick any two points on $\widetilde{\mu}(n)$ and join them with a two-parameter 
exponential fitting curve 
$g_1(n)=M\Lambda^{n},$
where $\Lambda$ is a coarse-graining of the true eigenvalues, then the curve $g_1(n)$ will 
always be above $\widetilde{\mu}(n)$ when it is between the two fitting points and below 
$\widetilde{\mu}(n)$ outside that interval. 
If we choose our fitting points to be the values of $\mu_2$ and $\mu_4,$ this gives us a simple 
lower bound for the negativity using only four replicas. 
There are systems for which this lower bound would be exact \cite{TopoRenyiDensity} 
but in general it will not be very accurate.
However it is possible to use the convexity of this problem to obtain tighter lower bounds using 
more moments. 

In this paper we will develop methods for bounding the set of admissable values for the 
unknown first moment $||\rho^{T_2}||_1$ based on those for studying a class of problems 
related to Laplace transform inversion, known as Hausdorff moment problems \cite{Hausdorff}. 
The moments $\{c_k\}_{k=0}^{n}$ are defined in terms of an integral over a finite range 
\cite{Akhiezer,ShohatTamarkin}
\begin{equation}\label{classicalForm}
 c_k = \int_a^b \lambda^k \sigma(\lambda) d\lambda.
\end{equation}
where the distribution $\sigma(\lambda)$ must be a non-negative function. 
If we only know a finite number of moments, we are dealing with an instance of the truncated 
moment problem.
The precise statement of the truncated moment problem is: Given a set of moments 
$\{c_k\}_{k=0}^n,$ 
find (one or more) distributions $\sigma(\lambda)$ consistent with the given moment values.
These integrals can be replaced by sums not only for finite dimensional systems, but also for extremal solutions of continuous systems, which will allow us to extremize integrals of the 
form
\begin{equation}\label{MarkovProblem}
I = \int_a^b \Omega(\lambda)\sigma(\lambda)d\lambda,
\end{equation}
where we have only incomplete knowledge of the distribution $\sigma(\lambda).$

\subsection{Properties of the set of possible solutions}

It is known \cite{KreinNudel} that the extreme points of the set of solutions 
(known as the \emph{canonical representations} of the moment system $\{c_k\}_{k=0}^{n}$) 
are distributions with a minimal number of point masses.  This set of representations have 
their own extrema, known as the \emph{principal representations}. 
We will refer to the positions of the weights as the ``roots'' of the representation, following 
\cite{KarlinStudden}. 

All other solutions lie in the convex hull of the boundary points \cite{KarlinShapleyOpus}.
As this is a continuously parametrized set infinite combinations are possible, such as those 
obtained by maximum entropy methods. 
The lower bounds found in this paper reflect the fact that the utility of a given supply of 
entanglement to a set of agents is limited by their knowledge of how much entanglement they have. 
If the parties to a distillation attempt are only able to perform operations on a small number 
of copies of a state at any given time, they may be unable to detect the presence of 
some entanglement and therefore be unable to distill it; there are states for which many 
copies are required for any distillation to succeed \cite{WatrousManyCopies}. 

It is also important to note that the positions of the point masses in the canonical 
representations are not arbitrary; one cannot simply pick $\kappa$ roots and then tune the 
weights at those positions to get a solution consistent with the moment constraints.  
Instead, the roots of the canonical representations are continuous functions of each other: 
move one, and the others adjust their positions to maintain consistency with the moment 
constraints.

It can be shown (subject to some conditions on $\Omega(\lambda),$ which include continuity 
and strict convexity over the range of integration \cite{KreinNudel,KarlinStudden})
that the set of distributions giving the extrema of the integral in \eqref{MarkovProblem} 
will always include a pair of principal representations \cite{KarlinStudden} unless the 
solution is unique, whereupon the two principal representations coincide. 
The \emph{lower principal representation} has that name because it achieves the lower bound 
for the simplest versions of the extremization problem in \eqref{MarkovProblem} and many 
other versions of this problem. 
(The other principal representation is known as the \emph{upper principal representation} 
because it usually corresponds to the upper bounds in such problems.)

Some of the conditions on $\Omega(\lambda)$ refer to the existence conditions for the moment 
problem to be soluble, so we will return to them after we have introduced the existence 
conditions below. 
It should be noted that it is possible to relax some of the conditions on $\Omega(\lambda)$  
in certain special cases and still be able to find the extrema using the principal 
representations, though this must be done with care and the methods tend not to generalize. 
However, some of the more straightforward generalizations that can be made have the effect 
of reversing the roles of the extremal representations, so that the lower principal 
representation can correspond to the upper bound, and vice versa\cite{KarlinStudden}.

\subsection{Existence conditions}\label{Econds} 

The precise form of the conditions for the Hausdorff moment problem to have a solution 
depend on whether we have an odd or an even number of moments available 
\cite{ShohatTamarkin,KreinNudel,Akhiezer}
because the extremal solutions are defined by an even number of parameters. 
Thus if we know an even number of moments the problem will be well-constrained, 
and if not it will be ill-constrained.

For the well-constrained case when we are given an even number of moments $\{c_k\}_0^{2m+1}$ 
a necessary and sufficient condition for a solution to exist is that the determinants of the 
Hankel matrices
\begin{equation}\label{WellHankela}
 H_1(k)= \sum_{i,j=0}^{m} (c_{i+j+1}-a c_{i+j})|i\rangle \langle j|
\end{equation}
and
\begin{equation}\label{WellHankelb}
 H_2(k)= \sum_{i,j=0}^{m} (b c_{i+j}-c_{i+j+1})|i\rangle \langle j| 
\end{equation}
must be greater than or equal to zero for all $k \leq m.$ 
If at least one of these inequalites is zero, the system is singularly positive and there is 
only one solution. 
If all the inequalities are strict, then the system is called \emph{strictly positive}.

If instead we have an odd number of moments $\{c_k\}_{k=0}^{2m}$ we have an ill-constrained 
instance. Then the determinants of the set of Hankel matrices 
\begin{equation}\label{BadHankel1}
 H_1(k)= \sum_{i,j=0}^{m} c_{i+j}|i\rangle \langle j|
\end{equation}
must be non-negative for all $k\leq m,$ and the determinants of 
\begin{equation}\label{BadHankelab}
 H_2(k)= \sum_{i,j=0}^{m-1} ((a+b)c_{i+j+1}-abc_{i+j}-c_{i+j+2})|i\rangle \langle j|
\end{equation}
for $k=0,\ldots, m-1$ must also be greater than or equal to zero.

If we have a strictly positive instance, there will be infinitely many distributions that are 
consistent with the set of known moments $\{c_k\}_{k=0}^{n}.$ 
If we have an even number $2\kappa$ of constraints, the canonical representations are 
those which consist of no more than $\kappa+1$ roots, with the exception of one with 
$\kappa$ roots, which will be the lower principal representation. 
If there are $2\kappa+1$ moments, the canonical representations are the those with $\kappa+1$ 
roots. 
For these systems the lower principal representation has $\kappa+1$ roots, and occurs when one 
of the roots reaches the lower integration endpoint, $a.$ 
The other $\kappa$ roots occur within the interior of the integration range $[a,b].$
Thus the well-constrained lower principal representations tend to collect its roots in the 
interior of the range of integration, whereas the ill-constrained version tends to push as 
much spectral weight as it can towards the lower end of the range of integration. 

In the event that the instance is singularly positive, there is a loss of independence in the 
constraint equations and the unique solution will have fewer point masses, mimicking one with 
fewer moments.
However, only instances with extremely simple spectra can be singularly positive when 
constrained by only a small number of moments: the eigenvalues of such a spectrum can only 
take at most $\kappa$ distinct values. 
More typical systems, in which the eigenvalues can take many different values or even feature 
quasi-continuous bands will have true spectra that are safely inside the interior of the set 
of all possible distributions consistent with the known moments. 
Thus if a given instance outright fails to satisfy the corresponding existence conditions, we 
should interpret this as meaning that the data from at least the highest order moment involved 
is too noisy to use.  If this happens we can progressively omit the highest order moments until 
we are left with a system of moments that admit a solution.

\subsection{Characteristic equations for the roots of the lower principal representations}\label{chareqns}

For $n=2m-1,$ the roots of the lower principal representation are the values of $\lambda$ 
which solve the equations \cite{KreinNudel}:
\begin{equation}\label{wellCharEqn}
 \det
 \begin{pmatrix} c_0 & c_1 & \ldots & c_{m-1} & 1 \\
                 c_1 & c_2 & \ldots & c_{m} & \lambda \\
                 \vdots & \vdots & \ddots & \vdots & \vdots \\
                 c_{m} & c_{m+1} & \ldots & c_{2m-1} & \lambda^{m} 
                 \end{pmatrix}
   =0.
\end{equation}
Likewise for $n=2m,$ the roots for the lower principal representation are the solutions of
\begin{equation}\label{illCharEqn}
 (\lambda-a)\det
 \begin{pmatrix} c_1-ac_0 & \ldots & c_{m}-ac_{m-1} & 1 \\
                 c_2-ac_1 & \ldots & c_{m+1}-ac_{m} & \lambda \\
                 \vdots & \ddots & \vdots & \vdots \\
                 c_{m+1}-ac_{m} & \ldots & c_{2m}-ac_{2m-1} & \lambda^{m} 
                 \end{pmatrix}
   =0.
\end{equation}

We will call the multiplicities of the eigenvalues $M_i.$ Then we can write
\begin{equation}\label{muSetupGen}
 c_n = \sum_i M_i \lambda_i^n = \int_{a}^{b} \lambda^{n-1}\rho^{T_2}d\lambda,
\end{equation}
for integer values of $n.$ The right hand side is almost of the form of a ``classical'' moment 
problem\cite{ShohatTamarkin}, in \eqref{classicalForm}. 
Since $\rho^{T_2}$ is not positive semi-definite it cannot play the role of the distribution 
$\sigma(\cdot),$ so we must choose an alternative. Note that we could define an alternative 
measure in eqn.\eqref{muSetupGen} by relabelling $n$ by a ``shift'' of $\tau,$ i.e., defining 
the distribution to be $\sigma_{\tau}(\lambda)=\lambda^{-\tau}(\rho^{T_2})^{1+\tau}.$

\subsection{Backwards extensions of moment systems}

It turns out that choosing $\tau=1$ and $\sigma(\lambda):=(\rho^{T_2})^2$ doesn't work because 
choosing $\Omega(\lambda):=1/\lambda$ leads to a singularity at $\lambda=0$ when we try to 
extremize the integral in \eqref{MarkovProblem}.  
Instead we will have to shift the zero-position of the problem in the opposite direction 
(i.e., choose $\tau=-1$) and define $\sigma(\lambda)=(\rho^{T_2})^{0},$ so that
\begin{equation}
 \mu_0 = \int_{a}^{b} \sigma(\lambda)d\lambda.
\end{equation}
The unknown zeroth moment $\mu_0$ is the number of non-zero eigenvalues. 
This means that the distribution $\sigma(\lambda)$ will be analogous to a state-counting 
function, modulo the fact that the eigenvectors in question were produced by the partial 
transpose operation, rather than being physical states.
However, we may not have any \emph{a priori} knowledge of the value of $\mu_0,$ apart from the 
dimension of the Hilbert space in which our problem lives. Even if we have restricted ourselves 
to some variational basis, we can't just use the dimension of that, as the partial transpose 
map may take the density matrix to something outside the original set of states.  

Therefore we will have to perform what is known as a \emph{backwards extension} 
(in the literature on the moment problem) 
to estimate $\mu_0,$ so we can then ``step forward'' to estimate $||\rho^{T_2}||_1$ by using 
$\Omega(\lambda):=\lambda.$ The existence conditions are necessary and sufficient for the 
moment problem to have at least one solution. We may therefore find a lower bound for $\mu_0$ 
by using those existence conditions.  

The need for a backwards extension is another reason for restricting this analysis to use only 
the lower principal representations, because the upper principal representations are generally 
not well-defined for backwards extensions. 
Finding expressions for them that appear to have the correct form is not difficult; 
the hard part of the problem is showing that they are informative estimates, or even that they 
correspond to the extrema at all. 

The fundamental reason why the problem of finding upper principal representations for backwards 
extensions is hard can be seen from the original intuition behind moment problems. Consider a rod 
on a pivot that has an unknown mass distribution along its length. 
Given only the moments of order two or higher, find the maximum possible mass of the rod. 
This is generically insoluble if there could be a finite mass accumulation at the pivot, because 
its presence cannot be detected by the other moments but it will still contribute to the total 
mass of the rod. 
Thus finding the upper principal representation of the mass distribution of the bar is impossible 
unless we can find the right kind of additional information about the problem. 
By contrast the lower principal representations require only a lower bound on $\mu_0,$ which can 
be obtained using the existence conditions, above. 

Once we have performed the backwards extension to lower bound $\mu_0,$ we can then proceed to 
solve the characteristic equations for the lower principal representations. Finding the weights 
for those roots is then a linear Gaussian elimination problem. 
The resulting estimates for $||\rho^{T_2}||_1$ are indeed extremal, though it turns out that some 
of them are actually upper bounds, in an example of the role-reversal phenomenon pointed out in 
ref.\cite{KarlinStudden}.

The final step is to perform the extremization in \eqref{MarkovProblem}, which requires that 
we revisit the conditions on $\Omega(\lambda)$ required for this method to work. This function 
is continuous, but it isn't strictly convex (only convex) so we need to check the more 
general conditions in \cite{KreinNudel,KarlinStudden}.
For our purposes, once we have chosen to back-step the problem, the conditions on 
$\Omega(\lambda)=\lambda$ will amount to ensuring that the determinant 
\begin{equation}\label{TsysCheck}
 \det
 \begin{pmatrix} 1 & \lambda_0^2 & \ldots & \lambda_0^n & \lambda_0 \\ 
                 1 & \lambda_1^2 & \ldots & \lambda_1^n & \lambda_1 \\ 
                 \vdots & \,     & \ddots & \,          & \vdots \\
                 1 & \lambda_{n+1}^2 & \ldots & \lambda_{n+1}^n & \lambda_{n+1}  
\end{pmatrix}  
\end{equation}
does not change sign for any pairwise distinct values of $\{\lambda_i\}_{i=0}^{n+1}$ 
in $[a,b].$ 
A cyclic permutation of the columns shows that this will revert to the standard existence 
criterion for the power moment problem, albeit with the sign reversed for one of the cases. 
(This will lead to a role-reversal between the upper and lower principal representations.)

\section{Obtaining the estimates}

The most general range of integration $[a,b]$ for this problem is between $a=-1/2$ and 
$b=1,$ as shown in \cite{computEnt}. 
However we will continue to write the integration limits as $a$ and $b,$ because there may 
be situations in which the range of the eigenvalues is known to be smaller and incorporating 
that information will yield tighter bounds. Thus we can write 
\begin{equation}\label{muSetupOdd}
 c_n = \sum_i M_i \lambda_i^n \int_{a}^{b} \lambda^{n-1}\rho^{T_2}d\lambda.
\end{equation}
In the context of an experiment when we might have only three replicas available, the only 
non-trivial existence condition is that $\nu_3-\mu_2^2\geq 0,$ since 
$\nu_1=$Tr$(\rho^{T_2})=1.$
Given that, we may proceed directly to obtaining a lower bounds for $\mu_0$ and hence for 
$\mu_1.$ 
Therefore we will exhibit the initial safety checks for four and five replicas to show the 
forms for even and odd numbers of replicas respectively. For the initial existence check 
the problem is in the well-constrained form if the largest number of replicas used is even. 
Thus for four replicas eqn.\eqref{WellHankela} becomes
\begin{equation}
 \det\begin{pmatrix} (\mu_2-a)     & (\nu_3-a\mu_2) \\
                     (\nu_3-a\nu_2) & (\mu_4-a\nu_3)
 \end{pmatrix}
 \geq 0,
\end{equation}
and eqn.\eqref{WellHankelb} becomes
\begin{equation}
 \det\begin{pmatrix} (b-\mu_2)      & (b\mu_2-\nu_3) \\
                     (b\mu_2-\nu_3) & (b\nu_3-\mu_4)
 \end{pmatrix}
 \geq 0.
\end{equation}

Conversely for five replicas we are initially dealing with the ill-constrained version of the 
problem, so we must have that eqn.\eqref{BadHankel1} becomes
\begin{equation}
 \det\begin{pmatrix} 1 & \mu_2 & \nu_3 \\
                     \mu_2 & \nu_3 & \mu_4 \\
                     \nu_3 & \mu_4 & \nu_5
 \end{pmatrix}
 \geq 0
\end{equation}
and eqn.\eqref{BadHankelab} is now
\begin{equation}
 \det\begin{pmatrix} (a+b)\mu_2 -ab -\nu_3     & (a+b)\nu_3-ab\mu_2-\mu_4 \\
                     (a+b)\nu_3-ab\mu_2-\mu_4  & (a+b)\mu_4+ab\nu_3-\nu_5
 \end{pmatrix}
 \geq 0,
\end{equation}
where we have used the fact that $\nu_1=1.$

We now perform the backwards extension, which has the effect of interchanging the 
well-constrained instances with the ill-constrained ones, as we have inserted an extra 
moment-like parameter into the problem. 
Rather than keep switching back and forth between the two types of solution, we will work 
through first the well-constrained case, and then the ill-constrained one.

\subsection{The well-constrained lower bounds}

This is the form to use if we start with data from $3,5,7\ldots$ replicas. 
We will use the three replica case as an example. The constraints for five replicas work the 
same way, only the matrices will be correspondingly larger.
There are two backstep constraints for the three replica case. The first one is 
\begin{equation}\label{lowerBnda}
 \det\begin{pmatrix} (1-a\mu_0) & (\mu_2-a)     \\
                     (\mu_2-a)  & (\nu_3-a\mu_2) 
 \end{pmatrix}
\end{equation}
and the second: 
\begin{equation}\label{lowerBndb}
 \det\begin{pmatrix} (b\mu_0-1) & (b-\mu_2)     \\
                     (b-\mu_2)  & (b\mu_2-\nu_3)
 \end{pmatrix}.
\end{equation}
The lower bound for $\mu_0$ is whichever one of these gives the higher value. 
The lower bound for $\mu_1$ can then be found by substituting that value for $\mu_0$ in the 
lower principal characteristic equations, and solving this to find the roots 
$\lambda_1,\lambda_2$ for the lower principal representation. 
We will follow the method in ref.~\cite{KreinNudel}. The relevant characteristic equation 
is \eqref{wellCharEqn}.
\begin{equation}
 \det\begin{pmatrix} 
            \mu_0 & 1 & 1 \\
            1 & \mu_2 & \lambda \\
            \mu_2 & \nu_3 & \lambda^2
 \end{pmatrix}
 =0.
\end{equation}
For larger systems these will need to be solved by numerical methods.
Now we substitute for $\mu_0$ with the lower bound we found above, 
to obtain the characteristic equation for the lower bound,
which can be solved using standard methods.
%
We will label the lower root $\lambda_1,$ which is the only one that can be negative when 
we have only three moments. If $\lambda_1 \geq 0$ then our lower bound for $\mu_1$ is zero. 
If $\lambda_1$ is negative, we need to find its weight $M_1.$ 
We can do this by Gaussian elimination, starting from
\begin{equation}
 c_j = \sum_i M_i \lambda_i^j
 \end{equation}
where $c_{2k}=\mu_{2k}$ and $c_{2k+1}=\nu_{2k+1}.$ In this lowest-order case $j=2,3$ only.
We do not need to find the weight of the positive eigenvalue, $M_2,$ as the negativity is the 
sum of the absolute values of the negative eigenvalues only. Thus we obtain a lower bound for 
the negativity from three replicas:
\begin{equation}\label{N3lower}
 \mathcal{N}_3 \geq \frac{\mu_2\lambda_2-\nu_3}{\lambda_1(\lambda_1-\lambda_2)},
\end{equation}
when $\lambda_1<0.$
For higher order systems where we have more moments to work with, the number of eigenvalues 
that may be less than zero is unknown; 
indeed it is possible for $\rho^{T_2}$ to have more negative eigenvalues than it does positive 
ones \cite{Rana,NatLargeNPT}, and still satisfy Tr$(\rho^{T_2})=1.$ Nevertheless, the basic 
principle remains that while we will need all of the eigenvalues to find the estimate, we will 
only need to find the weights of the negative eigenvalues: the weights of the positive 
eigenvalues can just be eliminated. Note that for strictly positive moment systems this 
estimate will almost invariably be an infimum instead of a minimum. 
This is because we have not invoked the fact that $\sigma(\lambda)$ is a counting function 
anywhere in the calculation above, so the weights have not been constrained to only take 
integer values. 
Therefore unless $M_1$ and $M_2$ happen to take integer values by coincidence, this lower 
bound cannot actually be obtained.

\subsection{The ill-constrained lower principal representation}

This is the form required if we start with data from $4,6,8,\ldots$ replicas. 
We will work through the four replica case as an example. The backwards extension constraints 
for $\mu_0$ are those in eqns.~\eqref{BadHankel1} and \eqref{BadHankelab}, starting from 
$i=j=0,$ so eqn.~\eqref{BadHankelab} becomes
\begin{equation}
 \det \begin{pmatrix}
               (a+b) -ab\mu_0 -\mu_2 & (a+b)\mu_2 -ab -\nu_3 \\
               (a+b)\mu_2 -ab -\nu_3 & (a+b)\nu_3 -ab \mu_2 -\mu_4
      \end{pmatrix}
 \geq 0.
\end{equation} 
Once again, the lower bound is given by the larger solution for $\mu_0.$
The lower principal characteristic equation \eqref{illCharEqn} for four replicas 
becomes
\begin{equation}
 (\lambda-a)\det\begin{pmatrix} (1-a\mu_0)     & (\mu_2-a)     & 1         \\
                                (\mu_2-a)     & (\nu_3-a\mu_2) & \lambda   \\
                                (\nu_3-a\mu_2) & (\mu_4-a\nu_3) & \lambda^2 
 \end{pmatrix}
 =0.
\end{equation}
This 
will have three roots, one of which will always be at $a,$ and the larger interior 
root $\lambda_2>0.$ Of the interior roots, only $\lambda_1$ may be of either sign. 
This type of lower principal representation accumulates as much weight as possible at the 
lower end of the range of integration. 
Recall that the negativity is the sum of the absolute values of the negative eigenvalues only.  
As such, a principal representation that tries to maximize the size and weights of those will 
actually give us an upper bound for the negativity, so the principal representations will swap 
roles in the ill-constrained cases.

Note that the estimate may be non-zero even if $\lambda_1>0$ because there must be a root 
at $a.$ However this does not necessarily mean that our estimate must be greater than or 
equal to $a,$ because we have not incorporated the constraint that the weights must take 
integer values. Instead the requirement that there be a root at $a$ even though its weight 
may be less than one indicates that our method will generally give a supremum rather than 
a maximum.

We can only omit finding weights for eigenvalues that we know are positive.  
Since we don't know the sign of $\lambda_1$ in general we will have to allow for both 
possibilities.
As before, $M_1$ and $M_2$ will be the weights for the eigenvalue with the same label, and 
we will label the weight of root at $a$ as $M_a.$ So our estimate can be written thus
\begin{equation}
 \mathcal{N} = \frac{1}{2}((|\lambda_1|-\lambda_1)M_1+M_a).
\end{equation}
We can now find the counting weights $M_1$ and $M_a$ by Gaussian elimination, starting from
\begin{equation}
 c_j = M_a a^j + \sum_i M_i \lambda_i^j
 \end{equation}
where $c_{2k}=\mu_{2k}$ and $c_{2k+1}=\nu_{2k+1}$ as before. In this case $j=0,\ldots,4.$
A little algebra gives us the following:
\begin{equation}
 M_a a=\frac{\lambda_1^2(\lambda_2-\mu_2)-\lambda_2\nu_3+\mu_4}{(a-\lambda_2)(a^2-\lambda_1^2)}
\end{equation}
and 
\begin{equation}
 M_1 \lambda_1=\frac{\lambda_2\nu_3-\mu_4-a^2(\lambda_2-\mu_2)}{(\lambda_2
                     -\lambda_1)(\lambda_1^2-a^2)}
\end{equation}
and therefore
\begin{multline}
 \mathcal{N}_4 \leq 
  \frac{\lambda_1^2(\lambda_2 -\mu_2)-\lambda_2\nu_3+\mu_4}{(a-\lambda_2)(a^2-\lambda_1^2)} \\
  +\left(\frac{|\lambda_1|-\lambda_1}{2\lambda_1}\right)
    \frac{\lambda_2\nu_3-\mu_4-a^2(\lambda_2-\mu_2)}{(\lambda_2-\lambda_1)(\lambda_1^2-a^2)}.
\end{multline}

\section{Summary and open questions}\label{SummaryEtc}

In this paper we have shown how to obtain families of infima and suprema for the entanglement 
negativity, given only a few, low order moments of the partially transposed density matrix. 
Estimates for the entanglement negativity can be obtained even in the absence of information 
about the number of non-zero eigenvalues in the system's partially-transposed density matrix. 
A lower bound can be obtained with just three replicas, and an upper bound with four.
The methods in this paper can be extended further if more replicas are available, 
though the existence conditions will need to be checked at each order because the method 
becomes progressively more sensitive to statistical noise as the order of the approximation 
increases.

The method presented in this paper is fully model-independent. If model-specific information 
can be included in the calculation, then more structure can be resolved within the spectrum. 
After the first version of this manuscript appeared, an analytic derivation of the 
distribution of the eigenvalues of the partially transposed density matrix for one-dimensional 
conformal field theories was presented in \cite{N-Spectrum}, including detailed numerical 
results.  

There are still some open questions regarding the interpretation of the bounds. In particular, 
are the families of lower bounds in this paper entanglement monotones 
\cite{VidalMonotones,simplifyMonoConds,DiffCondMonotones}, 
and can they be implemented as quantum circuits\cite{BrunPolynomial} to measure entanglement, 
or even perform batch-size limited entanglement distillation? If an entanglement distillation 
protocol is optimal, its yield will be an entanglement monotone\cite{VidalMonotones}. 
However, the converse is not necessarily true: the entanglement of formation is a monotone but 
does not correspond to the yield of any distillation protocol, because not all entanglement can 
be distilled\cite{boundEntanglementH3}.


\begin{acknowledgments}
This work is supported in part by the M. Hildred Blewett Fellowship of the American 
Physical Society, www.aps.org.
I would also like to thank the Department of Physics and Astronomy at the University of 
Waterloo, The Institute for Quantum Computing for additional support, 
and The Perimeter Institute for Theoretical Physics for hospitality. 
I would also like to thank Roger Melko, Alioscia Hamma, William Donnelly, Todd Brun,  
Marco Piani and Vern Paulsen for some interesting discussions, 
and Yichen Huang for informative comments on the first version of this manuscript.  
\end{acknowledgments}




%

\end{document}